\documentclass[%
 reprint,
 amsmath,amssymb,
 aps,nofootinbib,
  twocolumn,
]{revtex4-1}
\usepackage[usenames, dvipsnames]{color}
\usepackage[autostyle]{csquotes}
\usepackage{hyperref}
\usepackage{color}
\usepackage{graphicx}
\usepackage{dcolumn}
\usepackage{bm}
\usepackage{enumerate}
\usepackage{ulem}

\definecolor{MyDarkBlue}{rgb}{0,0.1,0.7}
\hypersetup{colorlinks,breaklinks=true,
  urlcolor={blue},citecolor={MyDarkBlue},linkcolor={Blue}}

\begin{document}


\title{Gravity from the Determinant of the Energy-Momentum: \\ 
Astrophysical Implications
}
\author{Hemza Azri$^{1}$}
\email{hm.azri@gmail.com}
\author{Salah Nasri$^{1,2}$}%
\email{snasri@uaeu.ac.ae; salah.nasri@cern.ch}
\affiliation{$^{1}$%
 Department of Physics, United Arab Emirates University,
Al Ain 15551 Abu Dhabi, UAE
}%
\affiliation{$^{2}$International Center for Theoretical Physics, Trieste, Italy}%
\date{\today}
\begin{abstract}
Determinants of the second-rank tensors stand useful in forming generally invariant terms as in the case of the volume element of the gravitational actions. Here, we extend the action of the matter fields by an arbitrary function $f(\bm{D})$ of the determinants of their energy-momentum, and the metric, $\bm{D}=|\textbf{det}.T|/|\textbf{det}.g|$. We derive the gravitational field equations and examine the nonlinear terms induced by the determinant, specifically, the inverse of the energy-momentum tensor. We also show that these extensions require a nonzero stress-energy tensor for the vacuum. We propose a scale-free model, $f(\bm{D})=\lambda\bm{D}^{1/4}$, and show how it induces the familiar invariant terms formed by the trace of the energy-momentum tensor by expanding the action around the stress-energy of the vacuum. We study the hydrostatic equilibrium equations for a neutron star by providing relevant values of the dimensionless constant $\lambda$. We show that the differences from the predictions of general relativity, in the mass-radius relations, which are sensitive to the equations of state, are conspicuous for $\lambda \sim -10^{-2}$. We also show that the model does not affect the predictions on the primordial nucleosynthesis when it is applied to the early radiation era. This novel and unfamiliar type of gravity-matter coupling can lead to a rich phenomenology in gravitational physics.
\end{abstract}

\maketitle

\section{\label{sec:intro}{Introduction}}
\label{sec: introduction}

Despite the great success of General Relativity (GR) as a theory of gravity, it is still facing various challenges. Theoretical problems such as spacetime singularities and the incomplete description  of the theory itself at the quantum level, as well as from the observational data which brings to the forefront the problems of the dark sector, may be an indication to the  need for going beyond GR \cite{observations, ishak, review_on_mog}.

There have been many attempts to modify GR in the last few years. First of all, gravity, as presently understood, is identified by the curvature of spacetime  which in turn results from the distribution of matter fields via their \textit{energy-momentum} tensor $T_{\mu\nu}$. Therefore, there are ways to modify GR either from the pure geometric (curvature) or from the matter (energy-momentum) parts.

One line of thought seeks to modify the geometric sector solely and consider scalar curvature invariant terms added to the Einstein-Hilbert action. The resulting theories such as the famous $f(R)$ gravity have been extensively studied and considered for a possible explanation of various cosmological problems \cite{f(R)}. The other line of thought is to modify the matter sector. In this way, the extension largely follows from the \textit{energy-momentum tensor} of matter fields by adding a general function of its trace $T$ or its square $T_{\mu\nu}T^{\mu\nu}$.
Among these,  are the familiar $f(R, T, T_{\mu\nu}T^{\mu\nu}, \ldots)$ constructed from both spacetime curvature $R$ and the stress-energy momentum tensor \cite{f(T)}. 

The shared property of all these terms is that they are formed by contracting spacetime indices, a standard procedure that allows  the covariant character of the theories to manifest  both  generally and locally. However, it is known that the determinant of the spacetime metric tensor is also crucial for the overall actions to be generally invariant. 
It turns out that not only the metric but any tensor of second rank can be considered. Thereby, the determinant of rank-two tensors are supported by the principle of general invariance, and therefore are allowed in the generalized gravitational actions unless forbidden by a specific symmetry.   

For instance, the square-root of the determinant of the Ricci tensor (usually referred to Eddington action) leads to the same field equations which can be derived from the standard Einstein-Hilbert action. Interestingly, the Ricci-determinant can be combined with the metric-determinant to finally form a scalar that is also relevant to gravity \cite{ricci_determinant, padbanabham}. Motivated by these determinant-based constructions, it is worth considering the matter sector as well by establishing a new type of matter-geometry interaction through its stress-energy momentum tensor.

In this paper, we will consider the \textit{determinant of the energy-momentum} tensor $|\textbf{det}.T|$ for the possible extension of the matter sector. The common practice is to build the volume element from the metric density, $|\textbf{det}.g|^{1/2}$ as in GR, and therefore \textit{gravity from the determinant of the energy momentum} will be based on a general function $f(\bm{D})$ of the scalar $\bm{D}=|\textbf{det}.T|/|\textbf{det}.g|$.      

Besides the fundamental difference from the usual extensions, the presence of the determinant will induce the inverse of the energy-momentum of matter fields in the equations of motion. In the cosmological context where the sources are approximated by perfect fluids, this contribution will in turn induce the inverse of the equation of state of the species \cite{det_cosmology}. Generally speaking, the inverse of the energy-momentum tensor is defined only when the determinant does not vanish everywhere, and therefore one has to consider the nonzero vacuum energy (the cosmological constant) in the absence of matter fields. Thus, at large scales, the solutions of these models are not completely flat but (anti-) de Sitter spacetimes.    

After setting up the general framework, we propose a specific model in which the coupling constant is dimensionless, i.e. $f(\bm{D})=\lambda\, \bm{D}^{1/4}$. We adapt the resulting field equation for a static spherically symmetric background and study the corresponding stellar structure. In order to constrain our coupling constant, we apply the model to the famed mass-radius relation of a neutron star. The latter has been a viable accommodation for testing gravity models at the strong field regime. However, since the equation of state is not rigorously constrained, we will study our model through various, mainly four different equations of state. We will solve the stellar structure equations numerically, and based on the resulting mass-radius relations we show how the predictions of our model become distinguishable from those of GR. This obviously depends on the strength of matter-gravity coupling induced by the determinant of the stress energy. Guided by the maximum measured mass and radius of the neutron star, we deduce the relevant constraints on the free parameter of the model. We will also highlight the possible effects on primordial nucleosynthesis.

The rest of the article is organized as follows: in section \ref{sec:general_action} we set up the general formalism starting from the action principle that includes the determinant of the energy-momentum. In section \ref{sec:scale_free_model}, we propose the scale-independent model and apply it to the perfect fluid sources. We then study the stellar structure in \ref{sec:stellar_structure} and conclude in section \ref{sec:conclusion}.

\section{Gravity from the determinant of the energy-momentum}
\label{sec:determinant}
\subsection{Gravitational action and the field equations}
\label{sec:general_action}
In this section, we will introduce our total gravitational action which  includes the standard GR action (Einstein-Hilbert) plus matter and extend it with  the determinant of the energy-momentum. We will derive the associated field equations by varying the action with respect to the main field (the metric tensor.) This can also be realized in the Palatini formalism if the GR action is written in terms of both metric and a symmetric connection as independent variables. In this paper, we will be interested only in the former. The invariant action reads
\begin{eqnarray}
\label{general_action}
S =&&\int d^{4}x \sqrt{|\textbf{det}.g|}
\,
\left\{
\frac{1}{2\kappa}\left(R-2\Lambda_{0}\right) + L^{\text{m}}[g]\right\} \nonumber \\
&&
+\int d^{4}x \sqrt{|\textbf{det}.g|}\,f(\bm{D}),
\end{eqnarray}
where $\kappa=8\pi G$, with $G$ being Newton's constant. In the first line of this action we have the Ricci scalar $R$, the cosmological constant $\Lambda_{0}$, and the Lagrangian density of matter fields $L^{\text{m}}[g]$. The latter describes the minimal coupling of matter to gravity, i.e. through only the metric. The novel quantity in the above action is $f(\bm{D})$ which is a general function of the scalar
\begin{eqnarray}
\label{D}
\bm{D} \equiv \frac{|\textbf{det}.T|}{|\textbf{det}.g|}~.
\end{eqnarray}
where $\textbf{det}.T$ is the determinant of the energy-momentum tensor $T_{\alpha\beta}$. This is given by its definition
\begin{eqnarray}
\label{det_t}
\textbf{det}.T=\frac{1}{4!}
\epsilon^{\alpha\beta\gamma\rho}\epsilon^{\bar{\alpha}\bar{\beta}\bar{\gamma}\bar{\rho}}
T_{\alpha\bar{\alpha}}T_{\beta\bar{\beta}}T_{\gamma\bar{\gamma}}T_{\rho\bar{\rho}},
\end{eqnarray}
where $\epsilon^{\alpha\beta\gamma\rho}$ is the anti-symmetric Levi-Civita (permutation) symbol. Like the determinant of any second rank tensor, the quantity (\ref{det_t}) transforms identically to $\textbf{det}.g$, thus, the quantity $\bm{D}$ transforms like a scalar and then the total action is invariant under general coordinate transformation. 

In varying the action, one has to take into account that the energy-momentum depends on the metric, hence, it is worth giving the expression for the variation of its determinant. This reads  
\begin{eqnarray}
\delta_{g}
\textbf{det}.T
=
\textbf{det}.T
\left(T^{\text{inv}}\right)^{\mu\nu}
\delta_{g}T_{\mu\nu},
\end{eqnarray}
where $\left(T^{\text{inv}}\right)^{\mu\nu}$ is the inverse of $T_{\mu\nu}$ such that $\left(T^{\text{inv}}\right)^{\nu\alpha}T_{\mu\alpha}=\delta^{\nu}_{\,\mu}$, and it is given by its definition as
\begin{eqnarray}
\label{inverse_t}
\left(
T^{\text{inv}}
\right)^{\mu\nu}
=\frac{1}{3!}
\frac{1}{\textbf{det}.T}
\epsilon^{\mu\alpha\beta\gamma}\epsilon^{\nu\bar{\alpha}\bar{\beta}\bar{\gamma}}
T_{\alpha\bar{\alpha}}T_{\beta\bar{\beta}}T_{\gamma\bar{\gamma}}.
\end{eqnarray}

Therefore, by varying the total action (\ref{general_action}), the gravitational field equations are obtained as
\begin{eqnarray}
\label{gravitational_equations}
G_{\mu\nu}=
 -\Lambda_{0} g_{\mu\nu}
+\kappa T_{\mu\nu} 
+\kappa f(\bm{D})g_{\mu\nu}
+2\kappa \bm{D}f^{\prime}(\bm{D})\mathcal{T}_{\mu\nu} \nonumber \\
\end{eqnarray}
where $G_{\mu\nu}$ is the Einstein tensor, $T_{\mu\nu}=L^{\text{m}}g_{\mu\nu}-2\delta L^{\text{m}}/\delta g^{\mu\nu}$ is the energy-momentum tensor of matter, and $f^{\prime}(\bm{D})=df/d\bm{D}$. The last term is given by
\begin{eqnarray}
\label{tau}
\mathcal{T}_{\mu\nu}=
&&
-g_{\mu\nu}
+L^{\text{m}}\left(T^{\text{inv}}_{\mu\nu} -\frac{1}{2}g_{\mu\nu}T^{\text{inv}} \right)
+\frac{1}{2}T^{\text{inv}} T_{\mu\nu} 
\nonumber \\
&&
+ 2 (T^{\text{inv}})^{\alpha\beta}\frac{\delta^{2}L^{\text{m}}}{\delta g^{\alpha\beta}\delta g^{\mu\nu}}.
\end{eqnarray}
where $T^{\text{inv}}$ is the trace of $(T^{\text{inv}})^{\mu\nu}$ (not to be confused with the inverse of the trace of $T_{\mu\nu}$), and $T^{\text{inv}}_{\mu\nu}=g_{\alpha\mu}g_{\beta\nu}\left(T^{\text{inv}}\right)^{\alpha\beta}$.

Therefore, invoking the determinant of the stress-energy tensor in the gravitational action modifies the Einstein's field equations as follows:
\begin{itemize}
    \item First, the geometrical part (curvature terms) is not altered since matter field are assumed to be coupled only to the metric. Then, the Bianchi identity, $\nabla^{\mu}G_{\mu\nu}=0$, can be easily applied. However, this will certainly lead to an extended energy-momentum conservation equation.
    \item An important contribution to the modified matter sector originates from the inverse of the energy-momentum tensor. Hence, the determinant of the energy-momentum tensor should not vanish. This is guaranteed if one considers the non-vanishing cosmological constant which is highly supported by observations. Indeed, the total energy-momentum tensor contains, in addition to matter fields' sources, the stress-energy tensor of the vacuum.
    \begin{eqnarray}
    T_{\mu\nu}^{(\text{tot})}= T_{\mu\nu}^{\text{(vac)}}+T_{\mu\nu}^{(\text{i})},
    \end{eqnarray}
    where $T_{\mu\nu}^{(\text{i})}$ is the energy-momentum tensor of all fluid types (matter and radiation) in the universe. The tensor $T_{\mu\nu}^{\text{(vac)}}=\mathcal{E}g_{\mu\nu}$ is the stress-energy of the vacuum where $\mathcal{E}$ being the vacuum energy density. The latter contains the bare cosmological constant $\Lambda_{0}$ and all non-vanishing contributions from the ground states of the quantum fields as well as from phase transitions (see Refs \cite{ccp} for detailed reviews). Cosmological observations imply a value around $10^{-10}\, \text{erg}/\text{cm}^{3}$ but not zero for the vacuum energy density. Thus, even in the absence of matter (and radiation) sources, i.e., when $T_{\mu\nu}^{(\text{i})}=0$, the energy-momentum tensor and its determinant do not vanish thanks to the cosmological constant term (the vacuum energy).   
\end{itemize}

\subsection{The case of perfect fluids}

Before proposing any particular model $f(\bm{D})$, one can calculate and examine the quantity $\bm{D}$ in terms of all sources of gravity at hand. If the latter are described by perfect fluids then one writes
\begin{eqnarray}
T_{\mu\nu}=\sum_{a} (\rho_{a} + P_{a})u_{\mu}u_{\nu}+\sum_{a}P_{a}g_{\mu\nu},
\end{eqnarray}
which contains all the content of the universe, namely, nonrelativistic matter and radiation as well as the vacuum (cosmological constant).

With all the contributions, one finds
\begin{eqnarray}
\bm{D}=\left| \left(\sum_{a}\rho_{a}\right)\left(\sum_{a}P_{a}\right)^{3}\right|.
\end{eqnarray}
We notice here that unlike the case of models extended by functions of the trace or the square of the energy-momentum, the contribution of the pressure controls this new quantity at the level of the action. However, it is worth noticing again that the inverse of both pressure and energy density will come out due to the emergence of the inverse of the energy-momentum after the variation of the action. 
That being said, we should emphasize again that, in general, the total energy-momentum tensor involves a nonzero vacuum energy. In fact, the last expression of the determinant can be written for matter and radiation ($a=\text{i}$), and the vacuum ($a=\text{vac}$) as  
\begin{eqnarray}
\label{D(rho,p)}
\bm{D}=
\left|
\left(\sum_{\text{i}}\rho_{\text{i}} +\mathcal{E}\right)
\left(\sum_{\text{i}}P_{\text{i}} -\mathcal{E}\right)^{3}
\right|.
\end{eqnarray}
Hence, for matter (or radiation) such as the case of the interior of massive objects, the cosmological constant is negligible, whilst in the vacuum (where $\rho_{\text{i}}$ and $P_{\text{i}}$ vanish), the cosmological constant becomes important, and in this case $\bm{D}= \mathcal{E}^{4}$ is nonzero.   

In the subsequent section we will propose a scale free-model; the first model of the energy-momentum-determinant gravity that we consider in practice. 

\subsection{Scale-free model}
\label{sec:scale_free_model}
As in extended gravity theories, one may think of $\bm{D}^{n}$ models where $n$ does not need to be an integer. In contrast, dimensional analysis favors an inverse of an integer. However, the high dimension of the determinant requires a high dimensional constant for the action to be dimensionless. With a mass scale $M$, this requires a general model of the form   
\begin{eqnarray}
f(\bm{D})=M^{4(1-4n)}\bm{D}^{n}.
\end{eqnarray}
Powers of $\bm{D}$ are not the only interesting models one may consider. In fact, from the field equations (\ref{gravitational_equations}) we notice the presence of the term $\bm{D}f^{\prime}(\bm{D})$ which can be absorbed by a logarithmic model, and leads to a simple dynamics. 

Nevertheless, as a first application of our setup based on the energy-momentum determinant, we will consider the scale-independent model, $n=1/4$ and $M^{0}\equiv \lambda$ (a dimensionless constant) which takes the form
\begin{eqnarray}
\label{the_model}
f(\bm{D})=\lambda \bm{D}^{1/4}.
\end{eqnarray}

Based on the generic case we studied above, the model (\ref{the_model}) is worth studying because it is the only resulting case that does not require a higher (mass) dimensional constant in the gravitational action (\ref{general_action}). Additionally, with the non-vanishing stress-energy of the vacuum (see the above discussion), this particular model can lead to interesting implications. In fact, the determinant structure (\ref{the_model}) can induce the familiar invariant terms formed by the trace of the energy-momentum tensor as an approximation. This can be realized if we expand the energy-momentum tensor around the stress-energy of the vacuum as
\begin{eqnarray}
T_{\mu\nu} \rightarrow
\mathcal{E}g_{\mu\nu} + T_{\mu\nu},
\end{eqnarray}
where $T_{\mu\nu}$ is a small perturbation. 

This leads to
\begin{eqnarray}
\textbf{det}.T
\simeq
\mathcal{E}^{4} \times
\textbf{det}.g
\times
\Bigg\{
1&&+
\frac{T_{\,\,\,\nu}^{\nu}}{\mathcal{E}}
+\left(
\frac{T_{\,\,\,\nu}^{\nu}}{2\mathcal{E}}
\right)^{2}
-\frac{T_{\,\,\,\beta}^{\nu} T_{\,\,\,\nu}^{\beta}}{2\mathcal{E}^{2}}
\nonumber \\
&& + \mathcal{O}\left(\frac{T_{\,\,\,\nu}^{\nu}}{\mathcal{E}}\right)^{3}
\Bigg\},
\end{eqnarray}
and finally
\begin{eqnarray}
\bm{D}^{1/4} \simeq
\mathcal{E}
&&
+\frac{1}{4}\,T
+\left(\frac{1}{16\mathcal{E}}\right)\,T^{2}
-\left(\frac{1}{8\mathcal{E}}\right)\,T_{\mu\nu}T^{\mu\nu} \nonumber \\
&&+\mathcal{O}(T/\mathcal{E})^{3}.
\end{eqnarray}
The first term in this expansion is the vacuum energy with a theoretical estimation of $\mathcal{E}\sim \Lambda_{\text{UV}}^{4}$, where $\Lambda_{\text{UV}}$ being an Ultra-Violet cutoff,  defined as the energy scale up to  which one trusts  quantum field theory. At the level of the action (\ref{general_action}), this term would simply contribute to the cosmological constant. 
The terms proportional to the trace of the energy-momentum tensor have been considered separately in the literature as possible models to be applied to cosmology and astrophysics, such as the $f(R,T)$ and energy-momentum squared gravity \cite{f(T)}. These models seem to arise from the determinant structure we proposed here.

Next, we will explore the astrophysical implication of this model where we will focus on the stellar structure for neutron stars. We will give a thorough phenomenological study and examine the new effects by putting relevant constraints on our free parameter.

\section{Astrophysical implication}

In this section we restrict the study to the interior of a neutron star where the cosmological term is ignored compared to matter. To simplify the calculation, we simply use $\rho_{\text{i}}=\rho$ and $P_{\text{i}}=P$ for the single fluid describing the star. Then, from (\ref{D(rho,p)}), one writes $\bm{D}=\rho P^{3}$ which does not vanish inside the star.    

In this case, the gravitational field equations (\ref{gravitational_equations}) read (taking $\kappa=1$)
\begin{eqnarray}
\label{eom_rho_p}
G_{\mu\nu}=&&
\left(\rho+P \right)u_{\mu}u_{\nu}
+Pg_{\mu\nu} \nonumber \\
&&
+\lambda \bm{D}^{1/4}
\left\{
g_{\mu\nu}
+\left(
1+\frac{P}{4\rho}
+\frac{3\rho}{4P}
\right)
u_{\mu}u_{\nu}
\right\}
\end{eqnarray}

However, as stated above, outside the star the energy-momentum tensor will be described in terms of the cosmological constant. In this vacuum case, the Lagrangian $L^{\text{vac}}=P^{\text{vac}}=-\mathcal{E}$ and the energy-momentum tensor $T^{\text{vac}}_{\mu\nu}=\mathcal{E}g_{\mu\nu}$. Hence, one can easily show that the gravitational action (\ref{general_action}) leads to the field equations in vacuum
\begin{eqnarray}
G_{\mu\nu}=-\Lambda_{\text{eff}}\,g_{\mu\nu},
\end{eqnarray}
where
\begin{eqnarray}
\label{effective_cc}
\Lambda_{\text{eff}}=
\Lambda_{0}+
\kappa(1-\lambda) \mathcal{E}
\end{eqnarray}
is simply a nonzero effective cosmological term. 

If the contribution to the vacuum energy comes only from the bare cosmological constant $\Lambda_{0}$, i.e., when $\mathcal{E}=\Lambda_{0}/\kappa$, one would simply have 
$\Lambda_{\text{eff}}=(1-\lambda)\Lambda_{0}$. Here, the factor of two in $\Lambda_{0}$ that would arise from the second term in (\ref{effective_cc}) is ignored because in this case the (similar) contributions from the Lagrangian $L^{\text{vac}}$ and the term $\Lambda_{0}$ in the action describe the same source.

The nonzero effective cosmological constant (\ref{effective_cc}) implies (anti-) de Sitter spacetime solution.

\subsection{Weak-field limit}
The weak field limit (or the Newtonian regime) of the previous equations will be derived now by considering small perturbations about all the quantities. We expand the energy density and pressure around their background values $\bar{\rho}$ and $\bar{P}$ as
\begin{eqnarray}
\rho \rightarrow \bar{\rho}+\rho, \quad
P\rightarrow \bar{P}+P,
\end{eqnarray}
and apply that for the perturbed spacetime metric in which the time-time component is given by the Newtonian potential $|\Phi| \ll 1$ as $g_{00}\simeq -1 -2\Phi$. 

In the weak field limit, one considers tiny stresses $T_{ij}$ (or pressure) compared to the energy density $T_{00}$, i.e. $|T_{ij}|/T_{00} \ll 1$. Therefore, we will neglect all the terms proportional to $P/\rho$ for both the background and first order terms. To that end, at first order, the time-time component of the gravitational equations (\ref{eom_rho_p}) reads
\begin{eqnarray}
\nabla^{2}\Phi=
4\pi G_{N}\left(
1+\frac{3\lambda}{4}\left(\frac{\bar{\rho}}{\bar{P}}\right)^{1/4}
\right)\rho
\end{eqnarray}
which describes the deviation from the standard Poisson equation when $\lambda \neq 0$. 

It is worth noting here that, unlike most of the familiar modified theories, this deviation is proportional to the inverse of the pressure which results from the inverse of the stress-energy tensor. This contribution is significant in the case of small pressure (compared to the energy density) as in the case of the Solar system which we consider here. 
The quantity $\bar{\rho}/\bar{P}$ in the second term can be constrained from the strength of the Newtonian gravitational potential of the Solar system which is nowhere larger than $10^{-5}$ \cite{will}. In fact, the pressure to which the system is subjected is comparable to $\rho |\Phi|$, hence, one has 
\begin{eqnarray}
\frac{\bar{P}}{\bar{\rho}} \sim |\Phi| \sim 10^{-5},
\end{eqnarray}
which leads to the modified Poisson equation
\begin{eqnarray}
\nabla^{2}\Phi \simeq
4\pi G_{N}\left(
1+0.13\times 10^{2} \lambda
\right)\rho.
\end{eqnarray}

In the following section we study the stellar structure equations for a neutron star based on the energy-momentum determinant gravity model.

\subsection{Stellar structure equations for a neutron star}
\label{sec:stellar_structure}
In what follows we apply our field equations (\ref{eom_rho_p}) to a static spherically symmetric background
\begin{equation}
ds^{2} = -{\rm e}^{2\Phi(r)}\, dt^{2} + {\rm e}^{2\Psi(r)}\,dr^{2}
+r^{2}\,d\theta^{2} + r^{2}\sin^{2}\theta \, d\phi^{2}
\end{equation}
where $\Phi(r)$ and $\Psi(r)$ are functions of the radial coordinate.

The following main equations can be easily derived by the same method used in GR \cite{mtw}. The main differences are in the right hand side of (\ref{eom_rho_p}) which includes the nonlinear terms in the energy density and pressure. Therefore, we find it unnecessary to put here all the details of the derivation.  

Using the field equations (\ref{eom_rho_p}), one finds the equation for the potential $\Phi(r)$ as
\begin{eqnarray}
\frac{d \Phi}{dr}=
\frac{m+4\pi r^{3}P}{r(r-2m)}
+\lambda
\bm{D}^{1/4}
\left(
\frac{4\pi r^{3}}{r(r-2m)}\right),
\end{eqnarray}
where we have introduced the mass $m(r)$ of the sphere of radius $r$.

As we have seen previously, outside the star (at large $r$), the spacetime approaches not the flat but (anti) de Sitter solution due to the presence of the cosmological constant. Hence, the exterior vacuum solution implies  
\begin{eqnarray}
{\rm e}^{2\Psi(r)}
=
\left(
1-\frac{2M}{r}+\frac{\Lambda_{\text{eff}}}{3}r^{2}\right)^{-1}.
\end{eqnarray}
Needless to say, although the term $\Lambda_{\text{eff}}$ is required in our theories in order to prevent the inverse of the energy-momentum from going singular, its effect however is purely cosmological, thus it must not affect the stellar structure.  

On the other hand, the radial component of (\ref{eom_rho_p}) gives the equation for the mass as
\begin{equation}
\label{eq_m}
\frac{d m}{dr}=
4\pi r^{2} \rho 
\left\{
1+
\frac{\lambda}{4}
\bm{D}^{1/4}
\left(
\frac{P}{\rho^{2}}
+\frac{1}{P}
\right)
\right\}.
\end{equation}

The extended TOV equation is derived by applying the Bianchi identity on (\ref{eom_rho_p}). This in turn leads to a generalized continuity equation from which one gets the equation for the pressure
\begin{eqnarray}
\label{tov}
\frac{d P}{dr}
=
-
&&
\left\{
\frac{m+4\pi r^{3}P}{r(r-2m)}
+
\lambda \bm{D}^{1/4}
\frac{4\pi r^{3}}{r(r-2m)}
\right\}(\rho+P)
\nonumber \\
&&
\times
\left\{
1
+
\frac{\lambda}{4}\bm{D}^{1/4} \left(\frac{1}{\rho} +\frac{3}{P}\right)
\right\} 
\nonumber \\
&& \times
\left\{
1+
\frac{\lambda}{4}\bm{D}^{1/4}
\left(
\frac{1}{c_{s}^{2}\rho}
+\frac{3}{P}
\right)
\right\}^{-1},
\end{eqnarray}
where $c_{s}^{2}=dP/d \rho$ is the speed of the sound wave.

The set of equations (\ref{eq_m})-(\ref{tov}) describes the stellar structure where their GR-limit is clearly understood for $\lambda=0$. We notice that the new contributions, induced by the determinant of the stress-energy of matter, are nonlinear in the energy density and pressure. Thus, analytical expressions of these quantities cannot be found easily. This necessitates a numerical solution which we present below.

\subsection{Constraints from neutron stars}
\label{sec:Neutron_stars:the_numerical_solution}
In this section we will put a constraint on the coupling constant $\lambda$ by studying the effects of the scale-free model (\ref{the_model}) on the mass-radius relation of the neutron stars. The latter have been used extensively in constraining various models of gravity in strong field regimes \cite{ns_constraints}.  

\begin{figure*}[t!]
\includegraphics[width=0.49\textwidth]{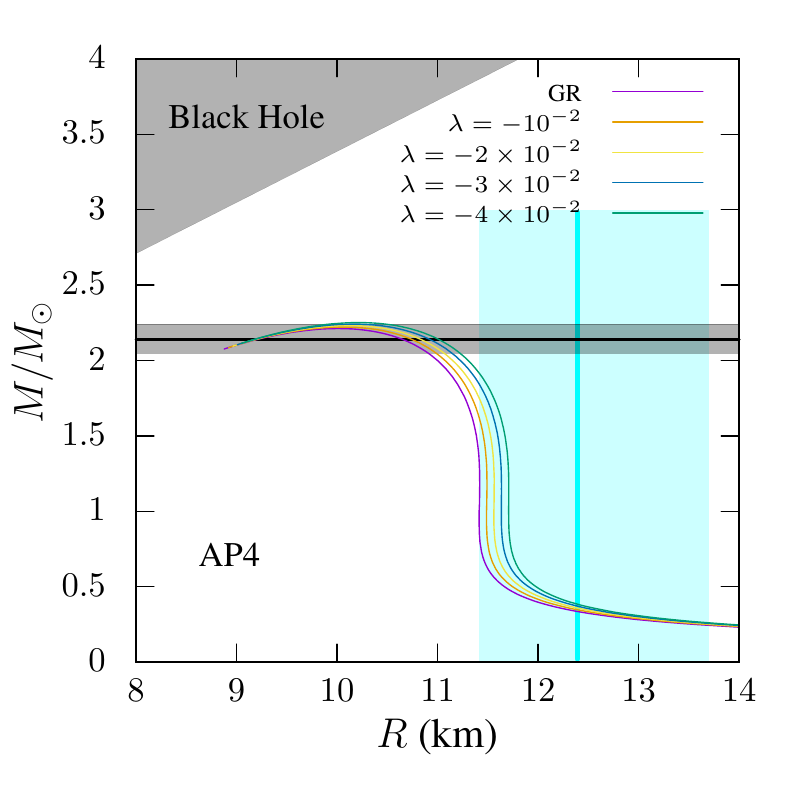}
\includegraphics[width=0.49\textwidth]{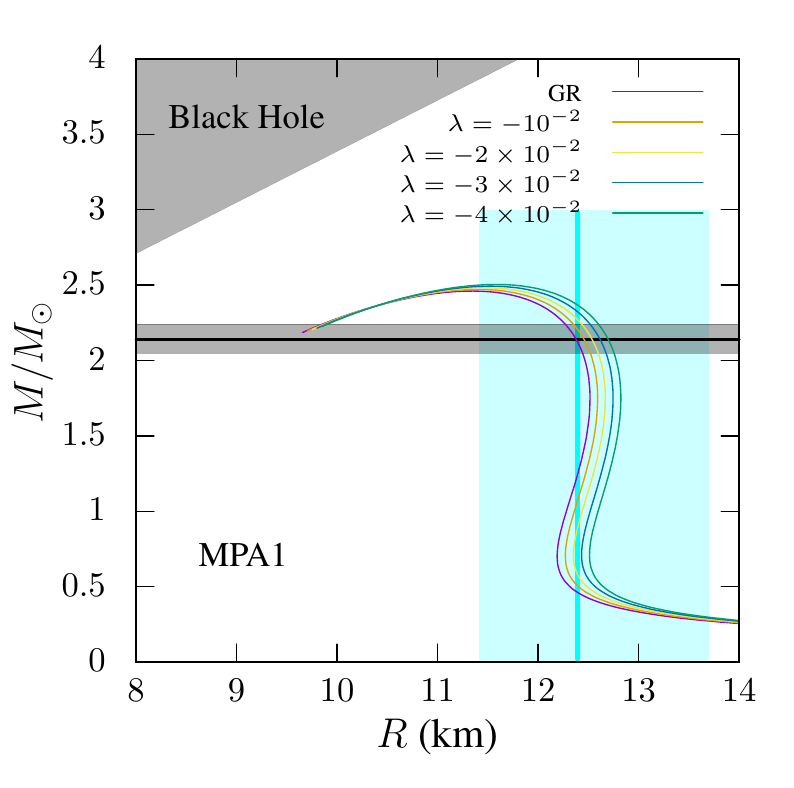} \\
\includegraphics[width=0.49\textwidth]{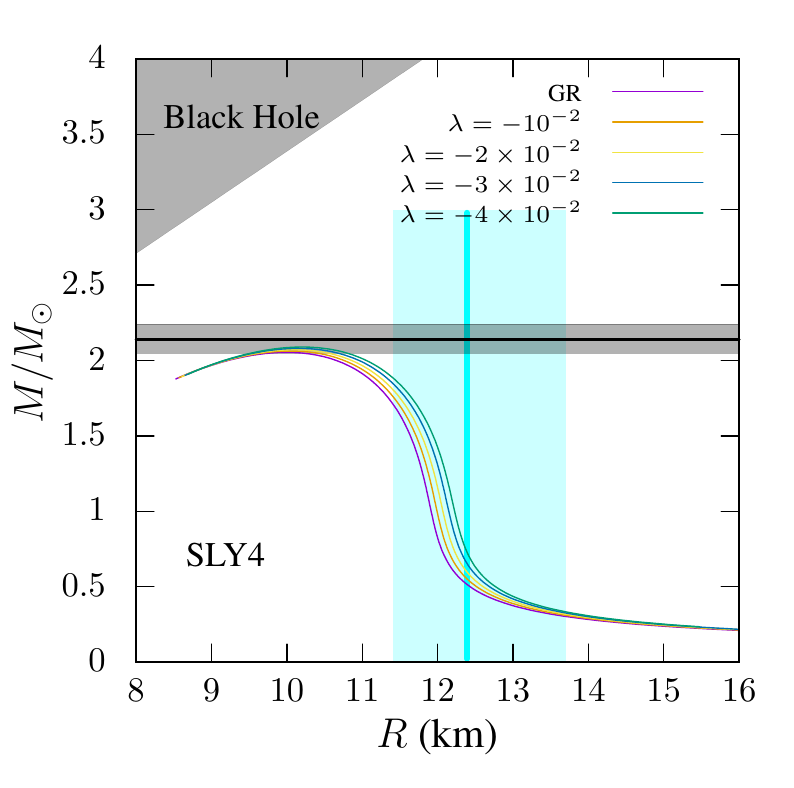}
\includegraphics[width=0.49\textwidth]{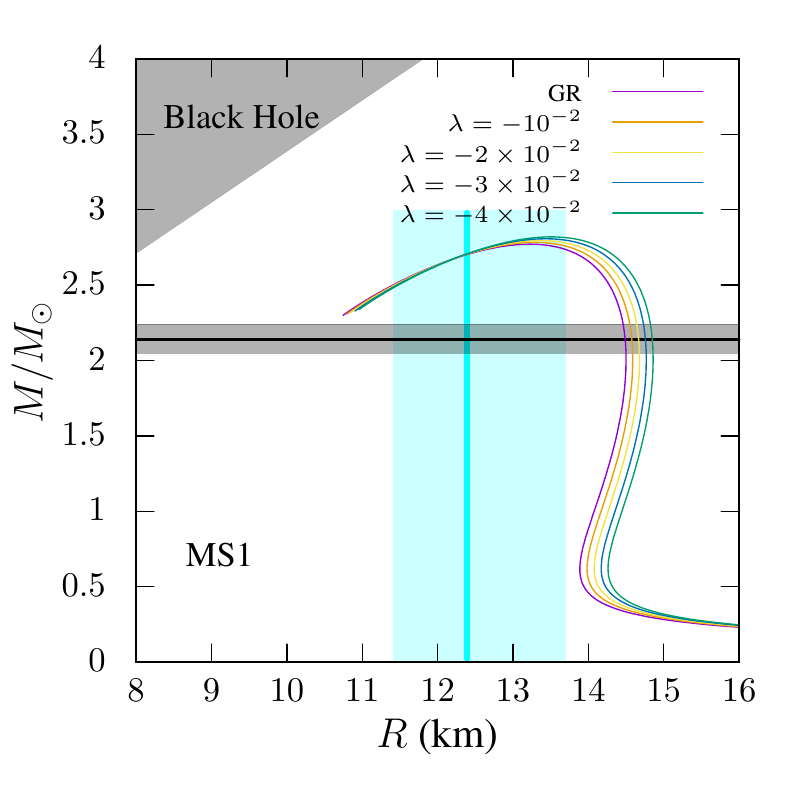}
\caption{Mass vs. radius of neutron stars from model (\ref{the_model}). Each panel corresponds to a different equation of state.
The gray shaded region corresponds to $R< 2GM/c^2$ (Black hole). The maximum measured mass $M = 2.14_{-0.09}^{+0.10}$ for a neutron star \cite{max_mr} is shown by the horizontal line.  The measured radius $R = 12.39_{-0.98}^{1.30}\,{\rm km}$ \cite{max_mr} is illustrated by the cyan colored shaded region. Departure from GR is clear for values around $\lambda \sim -10^{-2}$ for all the cases. However, the model is compatible with the maximum measured mass and radius in the case of MPA1.}
\label{fig_mr}
\end{figure*}

Here, we briefly describe the numerical method used to solve the hydrostatic equilibrium equations (\ref{eq_m})-(\ref{tov}). First, one has to close the system of equations by providing an equation of state relating pressure to the energy density, $P=P(\rho)$. In the case of neutron stars, the equations of state of the dense matter is not accurately constrained by the nucleon scattering experiments. Nevertheless, with various assumptions on the compositions and the nucleon-nucleon interactions, several equations of state are provided \cite{eos}. In our case, we will solve the main equations for four different equations of state, namely, AP4 \cite{AP4}, SLY4 \cite{SLY}, MPA1 \cite{MPA1}, and MS1 \cite{MS1}, by inspecting the effects of the novel contributions for each case.

For the numerical solution, we use the second order Runge-Kutta method (midpoint method). We also use the adaptive radial step-sizes \cite{BPS}
\begin{equation}
\Delta r = 0.01 \left(\frac{1}{m}\frac{dm}{dr} - \frac{1}{P}\frac{dP}{dr} \right)^{-1},
\end{equation}

which is adapted via the local mass and pressure gradients.

This technique permits for acceptable radial resolution at the regions of high pressure gradients. Furthermore, we keep the steps up to $10^3\,{\rm cm}$, but not larger.  
First, we choose a central pressure $P_{\rm c}$ (at the origin $r=0$) and $m(0)=0$ as a boundary condition for our system. We then integrate outwards up to the surface of the star where the pressure takes a very small value. We take for the surface pressure $P \sim 10^{10}\, \text{dyne}/\text{cm}^{2}$ which can be considered negligible compared to the pressure at the center, but yet is non-vanishing.

This point corresponds to the radius of the star $R$ and its total mass $M$. Last but not least, we obtain both mass and radius by varying the central pressure from $3\times 10^{33}$ to $9\times 10^{36}\,{\rm dyne\, cm^{-2}}$. Since our equations include a free parameter, we repeat the above process for various values of $\lambda$, and the above equations of state.

An important condition on $\lambda$ can be found from the requirement that the mass $m(r)$ should increase while integrating outwards starting from the center. Therefore, for general energy density and pressure, $dm/dr>0$ implies
\begin{eqnarray}
\left\{
1
-\frac{\rho +P}{2\rho}
\left(1 + \frac{3\rho -P}{2P}\right)
\right\}
\left(
\frac{P}{\rho}
\right)^{\frac{3}{4}}
\lambda
<1.
\end{eqnarray}

At the center of a typical neutron star one has $P/\rho \sim 0.2$ which leads to $\lambda > -0.8$. Therefore, the constraints on $\lambda$ from the measurements of the mass-radius relation (see below) must not fall below this negative value.

In Figure \ref{fig_mr}, we have depicted the mass of the neutron star versus its radius for both GR and the model (\ref{the_model}) for different values of $\lambda$ and for four different equations of state. In all four cases, we notice that the departure from GR becomes conspicuous starting from values $\lambda \lesssim -10^{-2}$. However, this deviation from GR falls within the region of the maximum measured mass and radius of a neutron star PSR J0740+6620, $M = 2.14_{-0.09}^{+0.10}$ and $R = 12.39_{-0.98}^{1.30}\,{\rm km}$ \cite{max_mr}, for the case MPA1.


\subsection{Effects on the big bang nucleosynthesis}
One of the most important predictions of the standard hot big bang cosmology is the determination of the abundances of the light elements formed in the early universe. Below we discuss briefly the possible deviations from the standard big bang-predicted abundances for $\mathrm{D}$ and $^{4}\mathrm{He}$. 

The possible deviations can be described by the expansion rate parameter $S\equiv H/H_{\text{SBBN}}$, where $H$ is the Hubble parameter associated to the present model and $H_{\text{SBBN}}$ for the standard big bang nucleosynthesis \cite{review_bbn}. These parameters are calculated during the early radiation era where the energy density of the universe is dominated by relativistic particles. 

From the gravitational field equations (\ref{eom_rho_p}), one can show that the energy density of the radiation evolves with the redshift $z$ as \cite{det_cosmology}
\begin{eqnarray}
\rho_{\text{r}}=\rho_{\text{r} 0}(1+z)^{\frac{4+10\times(1/3)^{3/4}\lambda}{1+7\times(1/3)^{7/4}\lambda}},
\end{eqnarray}
whereas the expansion rate is given by
\begin{eqnarray}
3H^{2}=
\kappa
\left(
1+\frac{7}{3}\left(\frac{1}{3} \right)^{4/3}\lambda
\right)\rho_{\text{r}}. 
\end{eqnarray}
The changes from the standard big bang nucleosynthesis arise when $S\neq 1$ where
\begin{eqnarray}
\label{S}
S=\frac{2\left(9+7\times 3^{1/4}\lambda\right)}
{3\left(6+5\times 3^{1/4}\lambda\right)}.
\end{eqnarray}

In \cite{constraint_on_s}, a limited range on the baryon density and the parameter $S$ are provided as $4\lesssim \eta_{10}\lesssim 8$ and $0.85 \lesssim S \lesssim 1.15$, respectively. Interestingly, the values of $\lambda$ predicted from the mass-radius relations of the neutron star (see Figure \ref{fig_mr} above) are compatible with this range. Hence, the scale-free model in (\ref{the_model}) does not alter the cosmological constraints on the abundances of the light elements although it leads to deviations from the standard hot big bang nucleosynthesis. We leave the thourough study of more cosmological implications of this model to a separate work \cite{det_cosmology}.

\section{Conclusion}
\label{sec:conclusion}
In this paper, we have considered a new type of matter-gravity coupling which is based on the determinant of the energy-momentum of matter fields. The main motivation is inherited from the fact that, like scalar invariant terms which are formed by contracting the covariant indices, the determinantal actions (mainly second rank tensors) also stands useful in curved spactimes.  

In a general framework, we have extended the standard matter Lagrangian by an arbitrary function $f(\bm{D})$ that involves the determinant of the energy-momentum and the metric so that the overall quantity, $\bm{D} \equiv |\textbf{det}.T|/|\textbf{det}.g|$, transforms as a scalar under the general coordinate transformations. We have derived the most general field equations from the action principle and examined the effects of the new terms induced by the determinant. Unlike the familiar extensions of gravity, we noticed the emergence of the inverse of the stress-energy tensor. We have discussed how the theory coincides with general relativity plus an effective cosmological constant in the case where the energy-momentum involves only the vacuum energy.  

To apply the work to an astrophysical phenomenon, we have proposed a model in which the free parameter is a dimensionless constant, i.e. a scale-independent model where $f(\bm{D})=\lambda \bm{D}^{1/4}$. When matter fields are approximated to perfect fluids, we have seen that, as expected, the novel contribution are nonlinear in the energy density and pressure, and involve their inverse. We derived the hydrostatic equilibrium equations and solved them numerically by assuming various equations of state for a neutron star. Then, we have presented our predictions in plots showing the important mass-radius relation for different values of the coupling constant $\lambda$. In this respect, we have focused on values that bring out noticeable departure from the predictions of general relativity, namely, values around $\lambda \sim - 10^{-2}$.    

We have also shed light on the possible effects of the scale-free model on the predictions of primordial nucleosynthesis. In this regard, we have shown that although the model must bring out slight deviations from the abundances of light elements predicted by the standard big bang nucleosynthesis, it still fits with observation. A separate work will be devoted to a more detailed study of the cosmological implications of this model \cite{det_cosmology}.  

Gravity models where matter fields enter the action principle through its determinant can lead to more interesting phenomenology. Therefore, various scale-dependent models different than the one presented here must be explored in various theoretical and phenomenological contexts. On the other hand, the rapid growth of the gravitational wave astronomy via LIGO, LISA, and other experiments is expected to probe precisely the gravitational physics \cite{gravitational_wave_astronomy}. This will provide a test for the validity of various extended theories of gravity including the stress-energy determinant gravity we explored in the present work \cite{gravitational_wave_astronomy_review}.

\section*{acknowledgments}
HA is indebted to Yavuz Ek\c{s}i for the help with the numerical solutions, and to Daniela Doneva for useful comments. He also thanks Glenn Starkman, Fabian Schmidt, and Albert Stebbins for fruitful discussions. This work is supported by the UAEU under UPAR Grant No. 12S004.

\end{document}